\documentclass[twocolumn,preprintnumbers,amsmath,amssymb]{revtex4}
\usepackage{graphicx}% Include figure files
\usepackage{dcolumn}% Align table columns on decimal point
\usepackage{bm}% bold math

\begin{document}

\preprint{APS/123-QED}

\title{Systematic study of the $^{87}$Sr clock transition in an optical lattice}

\author{Andrew D. Ludlow, Martin M. Boyd, T. Zelevinsky, Seth M. Foreman, Sebastian Blatt, Mark Notcutt, Tetsuya Ido, and Jun Ye }
 \affiliation{JILA and Department of Physics, National Institute of Standards and Technology and University of Colorado, Boulder, CO 80309-0440}

\date{\today}

\begin{abstract}
With ultracold $^{87}$Sr confined in a magic wavelength optical
lattice, we present the most precise study (2.8 Hz statistical
uncertainty) to-date of the $^1S_0$ - $^3P_0$ optical clock
transition with a detailed analysis of systematic shifts (20 Hz
uncertainty) in the absolute frequency measurement of 429 228 004
229 867 Hz. The high resolution permits an investigation of the
optical lattice motional sideband structure. The local oscillator
for this optical atomic clock is a stable diode laser with its
Hz-level linewidth characterized across the optical spectrum using a
femtosecond frequency comb.
\end{abstract}


\maketitle

In recent years, optical atomic clocks have become increasingly
competitive in performance with their microwave counterparts.  The
overall accuracy of single trapped ion based optical standards
closely approaches that of the state-of-the-art cesium fountain
standards \cite{Diddams1, Margolis1}. Large ensembles of ultracold
alkaline earth atoms have provided impressive clock stability for
short averaging times, surpassing that of single-ion based systems.
So far, interrogation of neutral atom based optical standards has
been carried out primarily in free space, unavoidably including
atomic motional effects that typically limit the overall system
accuracy \cite{Oates1, Sterr1, Ido1}.  An alternative approach is to
explore the ultranarrow optical transitions of atoms held in an
optical lattice \cite{Takamoto1,Porsev1,santra}. The atoms are
tightly localized so that Doppler and photon-recoil related effects
on the transition frequency are eliminated \cite{Ido2}. Meanwhile,
the trapping potential is created by laser light at a carefully
chosen wavelength ($\lambda_{\rm magic}$) such that it has
essentially no effect on the internal clock transition frequency.
Additionally, the increased atom-probe laser interaction time
enabled by the lattice confinement permits a full utilization of the
narrow natural linewidth. This optical lattice approach using
neutral atoms may provide the best possible combination of clock
stability and accuracy. Such a proposal has been under intensive
investigation in the case of the doubly forbidden $^1S_0$ - $^3P_0$
transition in the fermionic Sr isotope, $^{87}$Sr
\cite{Courtillot1,Takamoto1}. Similar work in Yb is also in progress
\cite{fortson,chad}.

The first high precision absolute frequency measurement was recently
reported for the $^{87}$Sr $^1S_0$ - $^3P_0$ clock transition using
a GPS-based frequency reference \cite{Takamoto1}. However,
establishing a new standard demands that different groups study
potential systematic errors. In this Letter, we present the most
precise study to-date of this ultranarrow clock transition in a
magic wavelength optical lattice, with a direct reference to the
NIST F1 Cs fountain clock. We have investigated systematic frequency
shifts including those originating from atomic density, wavelength
and intensity of the optical lattice, residual magnetic field, and
probing laser intensity. We determined the absolute frequency of the
$^{87}$Sr clock transition to be 429, 228, 004, 229, 867 $\pm$ 20
(sys) $\pm$ 2.8 (stat) Hz. Furthermore, we apply this optical clock
resolution towards a study of ultracold atomic motion in an
anharmonic optical potential.

To exploit the long coherence time provided by the ultranarrow
(transition linewidth $\gamma \simeq$ 1 mHz) $^1S_0$ - $^3P_0$ clock
transition, we have developed a cavity stabilized diode laser at 698
nm.  The laser is locked to an isolated, passive reference cavity
with finesse of $\sim$250,000 and cavity linewidth of $\sim$20 kHz.
This reference cavity is mounted vertically and supported
symmetrically with respect to the cavity length, which minimizes the
length sensitivity to environmental vibrations \cite{Notcutt1}. The
entire system is relatively simple and compact, occupying less than
1 m$^3$. To characterize this laser, we compare it to a highly
stabilized Nd:YAG CW laser (in an adjacent laboratory) which has
consistently demonstrated a sub-Hz linewidth. This comparison was
made possible with a phase-stabilized femtosecond frequency comb.
The repetition rate (100 MHz) of the octave spanning,
self-referenced Ti:Sapphire fs comb was tightly locked to the cavity
stabilized 698 nm diode laser, coherently transferring the diode
laser stability to each of the $\sim$10$^6$ modes of the fs
frequency comb.  A heterodyne beat signal between the sub-Hz Nd:YAG
reference laser and a corresponding fs comb mode at 1064 nm, shown
in Fig.\ 1, revealed a linewidth of less than 5 Hz for the 698 nm
diode laser and demonstrates optical coherence transfer at the
$10^{-14}$ level between remotely located lasers of different
colors. Additionally, the final stability of the 698 nm laser system
was confirmed to be limited by thermal-mechanical noise of the
mirror substrates in the Zerodur reference cavity \cite{Notcutt2}
and will be improved with new mirrors. The frequency drift caused by
the material creep of the cavity spacer was carefully compensated
via measurement of the optical frequency relative to a Cs
fountain-calibrated hydrogen maser using the same fs comb.

\begin{figure}[t]
\resizebox{8.5cm}{!}{
\includegraphics[angle=0]{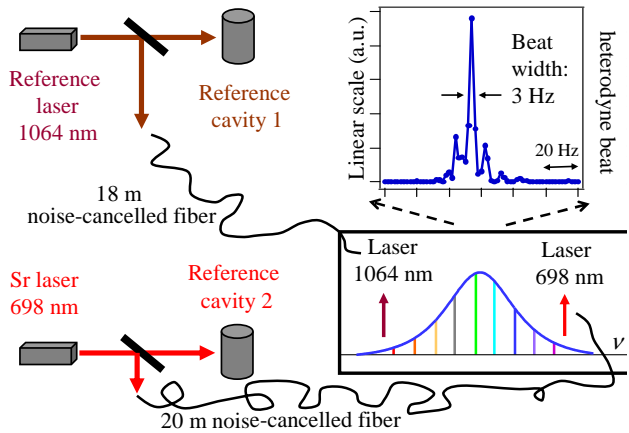}}
\caption{\label{Fig1}(color online) Measurement of probe laser
(698 nm) linewidth by comparison to a reference laser at 1064 nm
using an octave spanning fs laser. Linewidth of the heterodyne
beat at 1064 nm is 3 Hz.}
\end{figure}

Preceding the $^1S_0$ - $^3P_0$ spectroscopy, $^{87}$Sr atoms
undergo two stages of laser cooling (using the $^1S_0$ - $^1P_1$ and
then the $^1S_0$ - $^3P_1$ transitions - see Fig.\ 2) to prepare
approximately $10^6$ atoms at $\mu$K temperatures \cite{Loftus1,
Mukaiyama1, Xu1}. Loading of atoms from the MOT to a 1-D optical
lattice operating at $\lambda_{\rm magic}\approx$ 813 nm allows
simultaneous cooling and trapping on the $^1S_0$ - $^3P_1$
transition so that spatial mode matching between the MOT and lattice
is not critical. 10\% of the atoms ($\geq$ $10^5$) are loaded from
the red MOT to the lattice, which has a lifetime of 1 s.

Inside the lattice, the axial trap oscillation frequency is $\Omega
= 2\pi \times 80$ kHz, corresponding to a Lamb-Dicke parameter of
$\eta = k_p \sqrt{\hbar/(2m\Omega)} = 0.23$. Here $k_p =
2\pi/\lambda_p$, $\lambda_p$ is the probe wavelength, $2\pi\hbar$ is
Planck's constant, and $m$ is the atomic mass. In this regime,
spectroscopy of the clock transition can be performed nearly free of
any Doppler or photon recoil shifts. To avoid residual transverse
Doppler effects (transverse oscillation frequency is 500 Hz), the
probing laser is carefully aligned to co-propagate with the lattice
laser (Fig.\ 2), with both lasers being focused to a $\sim$70 $\mu$m
beam diameter at the trap. The probing laser is operated with
optical powers near 5 nW. Spectroscopy typically consists of a
probing time of 10-40 ms followed by a 2 ms illumination of the
atoms by 679 nm light, which pumps the excited state $^3P_0$
population to the $^3S_1$ state for shelving to the $^3P_2$ state
after spontaneous decay (Fig.\ 2). Because of spatial inhomogeneity
in the Rabi excitation frequency, we repeat the iteration of probe
and pump pulses approximately 20 times to enhance the signal size.
Finally, we detect the ground state $^1S_0$ population by resonantly
exciting the strong $^1S_0$ - $^1P_1$ transition at 461 nm and
counting fluorescence.

\begin{figure}[t]
\resizebox{8.5cm}{!}{
\includegraphics[angle=0]{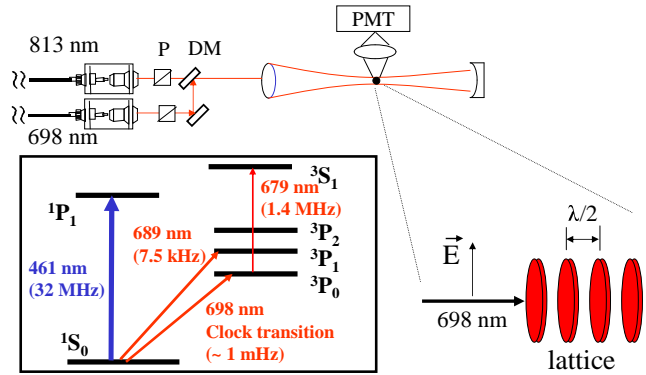}}
\caption{\label{Fig2}(color online) Partial diagram of $^{87}$Sr
energy levels and the schematic of the optical layout delivering the
probe and lattice laser to the atoms. P: polarizer, DM: dichroic
mirror, PMT: photo-multiplier tube. The strong longitudinal and weak
transverse confinement of the 1-D lattice creates 2-D trapping
regions.}
\end{figure}

The resolution of lattice spectroscopy enables us to investigate the
motional states of atomic Sr along the longitudinal axis of the 1-D
lattice. A deep lattice potential (where the populated motional
quantum states are well below the energy barrier between adjacent
sites) can be well approximated by harmonic confinement leading to
equally spaced motional states. These motional states allow
inelastic excitations which are red and blue detuned from the
elastic, purely electronic $^1S_0$ - $^3P_0$ excitation, with
amplitudes suppressed by $\eta^2/4$ in the Lamb-Dicke regime. A
measure of the amplitudes of the red and blue sidebands directly
gives the relative population of the ground and first excited
motional state in $^1S_0$ and thus the atomic ensemble temperature
\cite{Wineland1, Bergquist1}. As shown in Fig.\ 3(a), this measure
yields an atomic temperature of 5 $\mu$K, in good agreement with
time of flight temperature measurements. Since the lattice trap
depth is only $\sim$20 $\mu$K, the harmonic approximation
deteriorates for excited motional states ($n$ $\geq$ 1, where $n$ is
the motional quantum number). One consequence of the anharmonicity
of the optical confinement is the decreasing energy spacing between
different motional states, spreading the inelastic sidebands into
the individual $n \rightarrow n+1$ transitions. This anharmonic
spreading partly explains the broad sidebands shown in Fig.\ 3(a).
By knowing the trap depth and the energy of the $n$=0 $\rightarrow
n'$=1 transition, we solve for the exact eigenenergies of the
cos$^2(kz)$ dependent longitudinal lattice potential
\cite{Jauregui}, where $k$ is the lattice wave vector. The resulting
eigenenergies, in units of lattice photon recoil energy, are shown
in Fig. 3(b) as the lattice band structure. Knowing the relative
frequency of each $n \rightarrow n+1$ transition, we fit the blue
sideband of Fig.\ 3(a) with multiple Lorentzians to yield the
amplitudes (assumed to be Boltzmann-distributed) and the widths of
these transitions. The relative amplitudes among the various $n
\rightarrow n+1$ blue sidebands offer a more accurate measure of the
atomic population of various longitudinal motional states and hence
the sample temperature. This result, also 5 $\mu$K, is consistent
with measurements based on the relative blue and red sideband
amplitudes. The individual $n \rightarrow n+1$ transition linewidths
are very broad compared to the elastic $n \rightarrow n$ carrier
transition. Although some intrinsic broadening is introduced by the
band structure of the motional eigenstates ($<$1 Hz for lower
states, $\sim$kHz for upper states), major linewidth contributions
arise from externally induced dissipations among motional states and
radial variations of $\Omega$. Possible external dissipation
includes motional transitions induced by lattice phase and intensity
noise, two-body atomic collisions, and the blackbody radiation.
Unlike typical ion trapping experiments where individual motional
states are indistinguishable due to deep harmonic trapping, the
scenario described here permits detailed studies of level-specific
dissipation dynamics.

\begin{figure}[t]
\resizebox{8.5cm}{!}{
\includegraphics[angle=0]{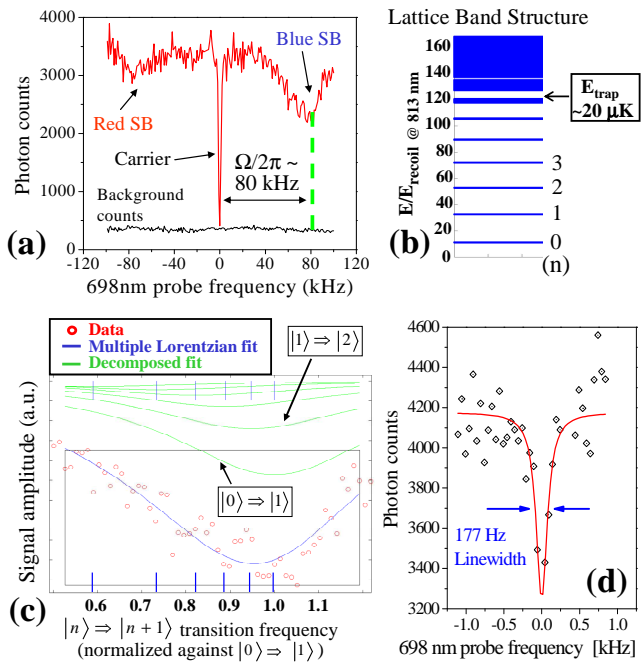}}
\caption{\label{Fig3}(color online) (a) Spectroscopic trace of
saturated, elastic $^1S_0$ - $^3P_0$ electronic transition and the
inelastic motional sidebands at 80 kHz detuning from the center
transition. (b) Lattice band structure derived from exact solution
of cos$^2(kz)$ dependent longitudinal lattice potential. (c) Fit to
the blue sideband of (a) where the sideband is composed of multiple
$n \rightarrow n+1$ transitions. Here the trap depth is $\sim$20
$\mu$K while the atomic sample temperature is 5 $\mu$K. The blue
vertical bars denote the position of each $n \rightarrow n+1$
transition from (b). (d) Narrow $^1S_0$ - $^3P_0$ elastic carrier
spectrum. }
\end{figure}

Spectroscopy of the $^1S_0$ - $^3P_0$ elastic carrier transition of
$^{87}$Sr in an optical lattice has the potential to yield one of
the highest optical resolutions ever measured \cite{Rafac1}. Of
course, a number of potential relaxation mechanisms can limit the
achievable coherence time.  A narrow spectrum of the carrier
transition is shown in Fig.\ 3(d). The $\leq$ 200 Hz full width at
half maximum linewidth allows for relatively rapid averaging of
multiple spectra to achieve statistical uncertainties of the
transition center frequency below the 5 Hz level, typically limited
by the hydrogen maser stability used for frequency counting. Our
efforts to reduce the broadening mechanisms were made in several
different areas. To eliminate linewidth broadening of the
fiber-optic-transferred probe laser ($\sim$100 Hz) caused by fiber
phase noise, we implemented fiber noise cancellation to sub-Hz
precision \cite{Ma1}. Lack of perfect overlap between the probe and
lattice lasers can introduce Doppler broadening along the weakly
confined transverse axes of the 1-D lattice. For this reason, we
ensured that the well overlapped lattice and probe lasers had strong
back coupling into both fibers from the retroreflecting lattice
mirror. Additionally, the finite size of the probe beam creates a
Heisenberg limited spread of $\overrightarrow{k_p}$ which samples
some of the transverse motion. This effect is below 50 Hz. In future
work, these broadening mechanisms will be circumvented by
implementation of a 3-D optical lattice.  State preparation of the
atomic population in ground state $m_F$ sublevels was not attempted,
making inhomogenous Zeeman broadening possible because of the
differential shift of $\sim$100 Hz/G $\times$ $m_F$ between $^1S_0$
and $^3P_0$ states due to the latter's hyperfine mixing
\cite{Peik1}. We used Helmholtz coil pairs to minimize residual
magnetic fields. If not operated at $\lambda_{\rm magic}$, the
lattice laser can introduce large inhomogenous AC Stark broadening.
By varying the lattice power and wavelength around $\lambda_{\rm
magic}$, we ensured linewidth contributions were below 30 Hz. Also,
dissipation processes in the higher motional states can cause
broadening in the elastic $n \rightarrow n$ transition. By sideband
cooling the population to the ground motional state, this effect
could be reduced \cite{Diedrich1}. To date, we have not observed any
effect of atomic density in the lattice on the measured linewidth.

\begin{table}[t]
\caption{Typical systematic corrections and their associated
uncertainties for the absolute frequency of the $^1S_0$ - $^3P_0$
clock transition.}
\begin{ruledtabular}
\begin{tabular}{lcc}
Contributor & Correction (Hz) & \hspace{0mm} Uncertainty (Hz) \\
\hline AC Stark shift (lattice) & -17 & 8.3 \\ AC Stark shift
(probe) & 7.0 & 0.9 \\ Blackbody shift & 2.2 & 0.02
\\ Zeeman shift & 0 & 12 \\ Recoil/Doppler shift & 0 & 1.9 \\ Density shift & -2.0 & 13 \\
Gravitational shift & -0.66 & 0.14 \\
Cs-maser calibration & 46.8 & 0.5
\\\hline Systematic total & 36 & 20 \\
\end{tabular}
\end{ruledtabular}
\end{table}

\begin{figure}[t]
\resizebox{8.5cm}{!}{
\includegraphics[angle=0]{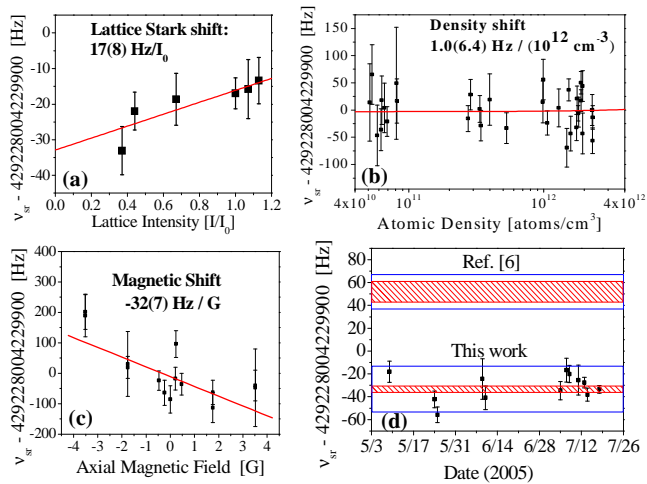}}
\caption{\label{Fig4}(color online) The measured $^1S_0$ - $^3P_0$
transition frequency versus (a) lattice intensity ($I_0$ = 35
kW/cm$^2$), (b) atomic density, and (c) magnetic field. (d) JILA
measurements over a 3 month period, with each data point
representing an averaged daily frequency measurement. The results
reported in this work (lower bars) and in Ref. \cite{Takamoto1}
(upper bars) are both shown with the total (outer box) and
statistical (inner shaded area) errors.}
\end{figure}

Figure 4 summarizes the measurement of the $^1S_0$ - $^3P_0$
transition frequency within three months. Limited by a statistical
uncertainty of 2.8 Hz, we have performed a detailed study of
systematic shifts and their corresponding uncertainties, which are
listed in Table 1. The most basic systematic error associated with
spectroscopy in an optical lattice is the AC Stark shift of the
clock transition due to the confining optical potential. To
experimentally determine the magnitude and uncertainty of this
shift, we varied the lattice intensity around our typical operating
intensity of $I_0$ = 35 kW/cm$^2$ at wavelengths below, above, and
near $\lambda_{\rm magic}$. At our typical operating wavelength of
813.437 nm, Fig. 4(a) shows the frequency shift as a function of
intensity. The slope of this shift yields an overall correction of
17(8.3) Hz at the typical lattice intensity of $I_0$. We have also
experimentally determined that for the lattice intensity of $I_0$,
the induced frequency shift is $\sim$2 mHz for a lattice frequency
deviation of 1 MHz from the magic wavelength. Combining this with
our measurement of the lattice AC stark shift of 17 Hz at 813.437 nm
yields $\lambda_{\rm magic}$ = 813.418(10) nm, in agreement with
\cite{Takamoto1}.

With potentially high atomic densities in an optical lattice,
characterization of density shifts is important. We have $\sim$400
lattice sites with typically $\sim$250 atoms per site, yielding
densities of $\sim$$10^{12}$ atoms/cm$^3$. Varying the atomic
density by a factor of 50 (Fig.\ 4(b)), we find the density shift at
our typical operating density of $2 \times 10^{12}$ atoms/cm$^3$ to
be 2(13) Hz. Asymmetric population distributions among $^1S_0$
ground state $m_F$ sublevels can lead to Zeeman shifts of the
transition frequency. We measured a shift of 32 Hz/G (Fig.\ 4(c)).
By keeping the magnetic field $<$400 mG during spectroscopy, the
Zeeman shift uncertainty is 12 Hz. The probe beam itself can induce
a frequency shift through two different physical mechanisms. The
first is the AC Stark shift of the $^1S_0$ and $^3P_0$ levels due to
their couplings to other states by the probe laser. For our
investigation, the AC Stark shift due to the probe laser was
exacerbated by using an electro-optic (EO) modulator to probe the
atoms with a weak sideband while retaining an off resonant carrier
for use in our fiber noise cancellation signal. The second frequency
shift mechanism arises from the posssibility of a small probe beam
misalignment with respect to the lattice laser, permitting photon
recoil shifts in the transverse direction of the optical trap. We
could separate these power-dependent effects by either varying the
total 698 nm light intensity incident on the atoms or changing the
relative carrier-sideband amplitude through the EO modulation index.
We determined each of these effects within 2 Hz uncertainty.

The frequency reference used for the optical measurement was a
hydrogen maser directly calibrated by the NIST F1 Cs fountain clock,
available to us by an optical fiber link from NIST to JILA
\cite{ye03}. The approximate 14 m lower elevation of our JILA Sr
experiment relative to the NIST fountain clock introduced a $<$ 1 Hz
gravitational shift. Including all systematic effects discussed
above, the total uncertainty (added in quadrature) is 20 Hz. The
final number we report for the $^{87}$Sr $^1S_0$ - $^3P_0$
transition frequency is 429, 228, 004, 229, 867 $\pm$ 20(sys) $\pm$
2.8(stat) Hz. We note this result disagrees by three standard
deviations with that of Ref. \cite{Takamoto1} measured with a
GPS-calibrated reference.

We have presented the most precise spectroscopic measurement of the
$^{87}$Sr $^1S_0$ - $^3P_0$ transition frequency in an optical
lattice with extensive studies of systematic uncertainties. We apply
this high resolution spectroscopy to investigating motional
properties of lattice trapped Sr atoms. This work demonstrates the
strength of the system for an optical atomic clock. Future work will
improve further upon the reported precision and accuracy.

We thank J. Hall, T. Loftus, C. Greene, and M. Holland for helpful
interactions and S. Diddams, T. Parker, and L. Hollberg for the
maser signal transfer. This work is funded by ONR, NSF, NASA, and
NIST.


\begin{thebibliography}{10}

\bibitem{Diddams1}
    S. A. Diddams {\it et al.}, Science {\bf 306}, 1318 (2004).

\bibitem{Margolis1}
    H. S. Margolis {\it et al.}, Science {\bf 306}, 1355 (2004).

\bibitem{Oates1}
    C. W. Oates, E. A. Curtis, and L. Hollberg, Opt. Lett. {\bf
    25} 21, 1603 (2000).

\bibitem{Sterr1}
    U. Sterr {\it et al.},  Comptes Rendus Physique {\bf 5}, 845 (2004).

\bibitem{Ido1}
    T. Ido {\it et al.}, Phys. Rev. Lett. {\bf 94}, 153001 (2005).

%\bibitem{Takamoto2}
%    M. Takamoto and H. Katori, Phys. Rev. Lett. {\bf 91}, 223001
%    (2003).

%\bibitem{Katori1}
 %   H. Katori in Proc. 6th Symp. on Freq. Standards and Metrology
 %   (ed. P. Gill), 323-330 (World Scientific, Singapore, 2002).

\bibitem{Takamoto1}
    M. Takamoto {\it et al.}, Nature {\bf 435}, 321
    (2005).

\bibitem{Porsev1}
    S. G. Porsev and A. Derevianko, Phys. Rev. A {\bf 69},
    042506 (2004).

\bibitem{santra}
    R. Santra {\it et al.}, Phys. Rev. A {\bf 69}, 042510 (2004).

\bibitem{Ido2} T. Ido and H. Katori, Phys. Rev. Lett. {\bf 91}, 053001
(2003).

\bibitem{Courtillot1}
    I. Courtillot {\it et al.}, Phys. Rev. A {\bf 68}, 030501(R)
    (2003).

\bibitem{fortson}
    T. Hong {\it et al.}, physics/0504216 (2005).

\bibitem{chad}
    C. W. Hoyt {\it et al.}, physics/0503240 (2005).

\bibitem{Notcutt1}
    M. Notcutt {\it et al.}, Opt. Lett. {\bf 30} 14, 1815 (2005).

\bibitem{Notcutt2}
    M. Notcutt {\it et al.}, Phys. Rev. Lett. (submitted).

\bibitem{Loftus1}
    T. H. Loftus {\it et al.}, Phys. Rev. Lett. {\bf 93}, 073003
    (2004); {\it ibid.} Phys. Rev. A {\bf 70}, 063413
    (2004).

\bibitem{Mukaiyama1}
    T. Mukaiyama {\it et al.}, Phys. Rev. Lett. {\bf 90}, 113002
    (2003).

\bibitem{Xu1}
    X. Xu {\it et al.}, Phys. Rev. Lett. {\bf 90}, 193002 (2003).

\bibitem{Wineland1}
    D. J. Wineland and W. M. Itano, Phys. Rev. A {\bf 20} 4,
    1521 (1979).

\bibitem{Bergquist1}
    J. C. Bergquist, Wayne M. Itano, and D. J. Wineland, Phys.
    Rev. A {\bf 36}, 428 (1987).

\bibitem{Jauregui}
    R. J{\'a}uregui {\it et al.}, Phys. Rev. A {\bf 64},
    033403 (2001).

\bibitem{Rafac1}
    R. J. Rafac {\it et al.}, Phys. Rev. Lett. {\bf 85},
    2462 (2001).

\bibitem{Ma1}
    L. S. Ma {\it et al.}, Opt. Lett. {\bf 19} 21, 1777
    (1994).

\bibitem{Peik1}
    E. Peik, G. Hollemann, and H. Walther, Phys. Rev. A {\bf 49},
    402 (1994).

\bibitem{Diedrich1}
    F. Diedrich {\it et al.}, Phys. Rev. Lett. {\bf 62},
    403 (1989).

\bibitem{ye03} J. Ye {\it et al.}, J. Opt. Soc. Am. B-Opt. Phys. {\bf 20}, 1459 (2003).


\end{thebibliography}
\end{document}